\documentclass[twocolumn,prl,floatfix]{revtex4}
\usepackage{graphicx}
\begin{document}
\title{Search for extraterrestrial point sources of neutrinos
with AMANDA-II}

\newcounter{foots}
\newcounter{notes}

\newcommand{\Aeff}[0]{A_\mathrm{eff}}
%
%
\author{
J.~Ahrens$^{11}$, 
X.~Bai$^{1}$, 
S.W.~Barwick$^{10}$, 
T.~Becka$^{11}$, 
J.K.~Becker$^{2}$, 
E.~Bernardini$^{4}$, 
D.~Bertrand$^{3}$, 
F.~Binon$^{3}$, 
A.~Biron$^{4}$, 
D.J.~Boersma$^{4}$
% ...
\vspace*{0.2cm}
}

\affiliation{$^1$Bartol Research Institute, University of Delaware, Newark, DE 19716}
\affiliation{$^2$Fachbereich 8 Physik, BU Wuppertal, D-42097 Wuppertal, Germany}
\affiliation{$^3$Universit\'e Libre de Bruxelles, Science Faculty CP230, Boulevard du Triomphe, B-1050 Brussels, Belgium}
\affiliation{$^4$DESY-Zeuthen, D-15735, Zeuthen, Germany}
\affiliation{$^5$Blackett Laboratory, Imperial College, London SW7 2BW, UK}
\affiliation{$^6$Dept. of Technology, Kalmar University, S-39182 Kalmar, Sweden}
\affiliation{$^7$Lawrence Berkeley National Laboratory, Berkeley, CA 94720, USA}
\affiliation{$^8$Dept. of Physics, Pennsylvania State University, University Park, PA 16802, USA}
\affiliation{$^9$Dept. of Physics, University of California, Berkeley, CA 94720, USA}
\affiliation{$^{10}$Dept. of Physics and Astronomy, University of California, Irvine, CA 92697, USA}
\affiliation{$^{11}$Institute of Physics, University of Mainz, Staudinger Weg 7, D-55099 Mainz, Germany}
\affiliation{$^{12}$Dept. of Physics, University of Maryland, College Park, MD 20742, USA}
\affiliation{$^{13}$University of Mons-Hainaut, 7000 Mons, Belgium}
\affiliation{$^{14}$Departamento de F\'{\i}sica, Universidad Sim\'on Bol\'{\i}var, Caracas, 1080, Venezuela}
\affiliation{$^{15}$Dept. of Physics, University of Wisconsin, Madison, WI 53706, USA}
\affiliation{$^{16}$Physics Dept., University of Wisconsin, River Falls, WI 54022, USA}
\affiliation{$^{17}$Division of High Energy Physics, Uppsala University, S-75121 Uppsala, Sweden}
\affiliation{$^{18}$Dept. of Physics, Stockholm University, SE-10691 Stockholm, Sweden}
\affiliation{$^{19}$Vrije Universiteit Brussel, Dienst ELEM, B-1050 Brussels, Belgium}


\begin{abstract}
We present the results of a search for point sources of high energy neutrinos in the northern
hemisphere using \mbox{AMANDA-II} data collected in the year 2000. 
Included are flux limits on several 
AGN blazars, microquasars, magnetars and other candidate neutrino sources. 
A search for excesses above a random background of cosmic-ray-induced 
atmospheric neutrinos and misreconstructed downgoing cosmic-ray muons reveals 
no statistically significant neutrino point sources.  We show that AMANDA-II has 
achieved the sensitivity required to probe known TeV $\gamma$-ray sources such 
as the blazar Markarian 501 in its 1997 flaring state at a level where neutrino and $\gamma$-ray fluxes are equal.

\end{abstract}

\maketitle

\section{Introduction}

The search for sources of high-energy extraterrestrial neutrinos is the primary
mission of the Antarctic Muon and Neutrino Detector Array (AMANDA). The mechanism for 
accelerating cosmic rays to energies 
above the ``knee'' ($10^{15} \, \mathrm{eV}$) remains a mystery.  Cosmic rays are thought to be 
accelerated in the shock fronts of galactic objects 
like supernova remnants, microquasars and magnetars, and in extragalactic sources such as  
the cores of active galaxies (AGN) and gamma ray bursts (GRB) \cite{halzen02}. 

High energy protons accelerated in these objects will collide with the ambient gas and
radiation surrounding the acceleration region, or with matter or radiation intervening between
the source and the Earth.
This leads to pion production, the charged pions decaying into 
highly energetic muon and electron neutrinos, and the neutral pions decaying into the observed $\gamma$-rays. 
Fermi acceleration of charged particles in magnetic shocks naturally leads to power-law 
spectra, $E^{\alpha}$, where $\alpha$ is typically close to -2. 
By the time the neutrinos reach the Earth, vacuum oscillations will have uniformly 
populated all three flavors (unless the neutrinos are unstable \cite{beacom}).   All limits quoted 
in this letter are on the muon neutrino flux arriving at the Earth; limits at the source will be approximately a factor of two higher due to oscillations.


Neutrino astronomy may provide information complementary to the knowledge
obtained from high energy photons and charged particles, since neutrinos
propagate directly from their point of origin undeflected by magnetic 
fields. Therefore they have the potential to reveal ``hidden'' sources masked
by photon absorption. Probing the neutrino sky may bring us closer to
solving the cosmic ray mystery, or might even reveal something completely new and unexpected.  
 
\section{AMANDA-II}

AMANDA-II \cite{nature} is a Cherenkov detector 
frozen into the antarctic polar icecap.  A high energy muon neutrino interacting with the 
ice or bedrock in the vicinity of the detector results in a high energy muon propagating
up to tens of kilometers.  

The muon track is reconstructed based on detection of the Cherenkov light emitted 
as it propagates through a 19-string array of 677 photomultiplier tubes.  
The median neutrino pointing resolution is 2 - 2.5$^\circ$, depending weakly on declination,
and is dominated by the resolution of the muon track reconstruction.  
AMANDA-II demonstrates a significant improvement over its predecessor in acceptance and 
background rejection, 
especially near the horizon.  
Results from the first phase of AMANDA, the 10-string subdetector AMANDA-B10,
have been reported in \cite{b10}.

Atmospheric muons from cosmic rays that penetrate to AMANDA depths are the dominant background. 
AMANDA-II views the neutrino sky above the northern hemisphere using the 
Earth as an atmospheric muon filter.  Cosmic rays also produce neutrinos in the Earth's atmosphere, 
but with a spectral index of $\alpha = -3.7$, softer than expected for astrophysical sources.
Atmospheric neutrinos are an important source for calibration in AMANDA \cite{amaatmos}, 
but are also background to a search for extraterrestrial point sources. A point source search is 
conducted by looking for excess events above the background, which is experimentally measured 
for a given angular search bin by taking the average background rate in the same band of 
declination.

\section{Data Processing \& Detector Simulation}
The data set comprises $1.2\times 10^{9}$ triggered events collected over 238 days 
between February and November, 2000, with 197.0 days livetime after correcting for 
17.2\% detector dead-time.
After application of an iterative series of maximum-likelihood 
reconstruction algorithms, 2.1 million events reconstructed with declination 
$\delta > 0^{\circ}$ remain in the experimental sample. 


To prevent bias in the selection of cuts, the data are ``blinded'' by randomizing 
the reconstructed Right Ascension (R.A.) angle of each event.     
The output of a Neural Network (NN) trained on simulated events and using six input 
variables (such as the number of unscattered photon hits, track length, likelihood of the muon track 
reconstruction, and topological variables \cite{amaatmos}), is used as a quality cut.
A second cut is placed on the likelihood ratio (LR) between the muon track reconstruction and 
a muon reconstruction weighted by an atmospheric muon prior \cite{bayesian}. The prior describes
the zenith-dependent frequency of downgoing muons such that choosing a cut on the LR gives 
downgoing hypotheses a prior weight of up to $10^{6}$ more than the upgoing hypotheses,
effectively forcing events surviving the cut to be of higher quality.  The final choice 
of NN quality cut, likelihood ratio cut, and the optimum size for a search bin are determined 
independently in each $5^\circ$ band of declination in order to optimize the
limit setting potential of the experiment \cite{mrp}.
The directional information is then restored (data ``unblinded'') for the calculation of the 
limits and significances. 

A full simulation chain \cite{b10} including neutrino absorption in the Earth, neutral current 
regeneration, muon propagation and detector response is used to simulate the point source signal
according to an $E^{-2}$ energy spectrum.  
The limits obtained in this analysis are a function of the measured background, $n_{b}$, 
as well as the expected number of events, $n_{s}$, from a simulated flux
$\Phi(E)$:  
$\Phi_{\mathrm{limit}}(E) = \Phi(E) \times \mu_{90}(n_{\mathrm{obs}}, n_{b})/n_{s}$ where $n_{\mathrm{obs}}$ is the 
number of observed events in the given source bin, and $\mu_{90}$ is the 90\% upper limit on the 
number of events following the unified ordering prescription of Feldman and Cousins \cite{fc98}.  

\section{Systematic Uncertainties}
   

Atmospheric neutrinos were used to determine the absolute normalization of the detector
simulation.  The normalization factor 0.86 is consistent with the theoretical uncertainty
of 25\% on the atmospheric neutrino flux \cite{amaatmos}.  The overall systematic 
uncertainty, which includes the theoretical uncertainty of the atmospheric neutrino flux
and statistical uncertainty of the measured background, is incorporated into the limits
using the Cousins-Highland \cite{ch92} prescription with unified Feldman-Cousins
ordering \cite{fc98,conrad} but with a more appropriate choice of the likelihood
ratio test \cite{hill03}.


Coincident events between the SPASE air shower array \cite{spase} and AMANDA-II were used to 
evaluate the systematic error in pointing accuracy.  This value was determined to be less than
$1^{\circ}$, which results in signal loss in a typical search bin of only 5\%.


\section{Results}

\begin{figure}[htb]
\begin{center}
\includegraphics[width=3.3in]{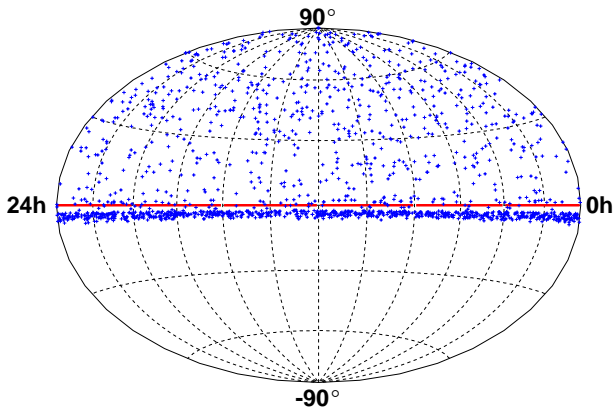}  
\end{center}
\caption{Final point source search sample plotted in equatorial coordinates. The thick band of events 
at $\delta < 0$ shows the onset of cosmic muon background contamination.}
\label{skyplot}
\end{figure}

The final sample consists of 699 upwardly reconstructed ($\delta > 0^{\circ}$) events, illustrated in 
Fig. \ref{skyplot}.
A comparison to the normalized atmospheric neutrino simulation reveals that for declinations 
$\delta > 5^{\circ}$ the sample is strongly dominated by atmospheric neutrinos.

A binned search for excesses in the region \mbox{$0^{\circ} < \delta < 85^{\circ}$} has been 
performed.  The search grid contains 301 rectangular bins with zenith-dependent widths ranging
from \mbox{$6^{\circ}$ to $10^{\circ}$,} based 
on the aforementioned bin-size optimization. The grid is shifted 4 times in
declination and right ascension to fully cover boundaries between the bins of the original 
configuration.  
The number of times to shift the grid was studied, taking into account statistical
penalties for the number of shifts by using simulated event samples, and set at a level where 
further shifts do not markedly improve the average maximum significance obtained on simulated 
Poisson-fluctuated signals of similar magnitude to the background. 

The most significant excess, observed 
at about $68^{\circ} \, \mathrm{Dec.,} \, 21.1 \mathrm{h} \, \mathrm{R.A.}$, is 8 events observed on an 
expected background of 2.1. Simulation reveals a probability of 51\% to
observe such an excess as a random upward fluctuation of
the background. 

\begin{figure}[htb]
\begin{center}
\includegraphics[width=3.3in]{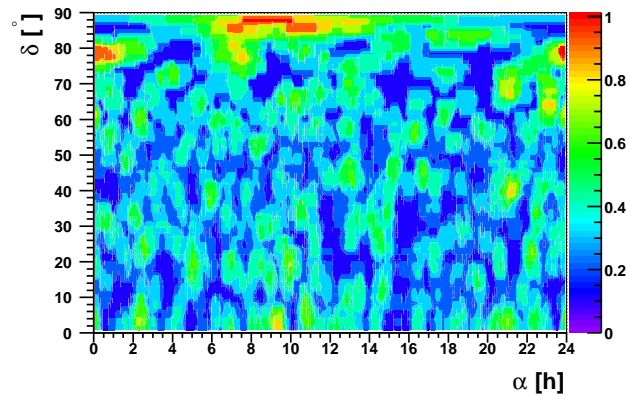}  
\end{center}
\caption{Neutrino flux upper limits (90\% confidence level) in equatorial 
coordinates. Limits (scale on right axis) are in units of 
$10^{-7} \mathrm{cm}^{-2} \, \mathrm{s}^{-1}$ for an assumed $E^{-2}$ spectrum, integrated above $E_{\nu} = 10\,\mathrm{GeV}$.  
Systematic uncertainties are not included.}
\label{limits}
\end{figure}

\begin{figure}[htb]
\begin{center}
\includegraphics[width=3.8in]{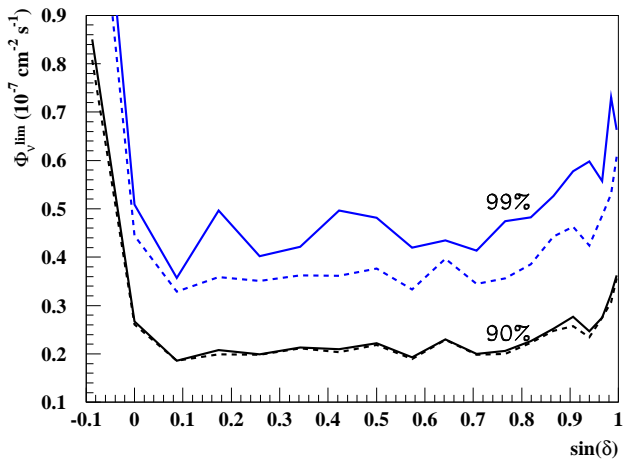}  
\end{center}
\caption{Average flux upper limits (90\% and 99\% confidence) for an $E^{-2}$ signal hypothesis, 
integrated above  $E_{\nu} = 10 \, \mathrm{GeV}$, are shown vs. declination ($\delta$), solid 
lines with, and dashed lines without the inclusion of systematic uncertainty (as described 
in the text).  The limits worsen near the horizon due to the onset of cosmic ray muon 
contamination.}
\label{sens}
\end{figure}

In addition to the binned search, we place limits on a number of extragalactic and galactic 
candidate sources. Circular bins with optimized radii are positioned at each
candidate position; the number of expected background events is given by the
number of observed events in the declination band scaled down to the search
bin area. The same method applied to any point in the Northern hemisphere
yields neutrino flux upper limits shown in Fig. \ref{limits}. For the region
$\delta>85^\circ$ an adjacent region at lower declination was used for the
background estimation.
The average flux upper limits 
for $E^{-2}$ spectra obtained in an ensemble of identical experiments in 
the case of no true signal is shown vs. declination in Fig. \ref{sens}. 
It should be pointed out that due to the large number of cells tested a single
source with a flux at the sensitivity level (90\% average flux upper limit) would normally be interpreted
as a statistical fluctuation. Therefore a flux a few times higher would be
needed to make a discovery of a point source possible in the binned search.

\begin{figure*}[htb]
\begin{center}
\includegraphics[width=3.5in]{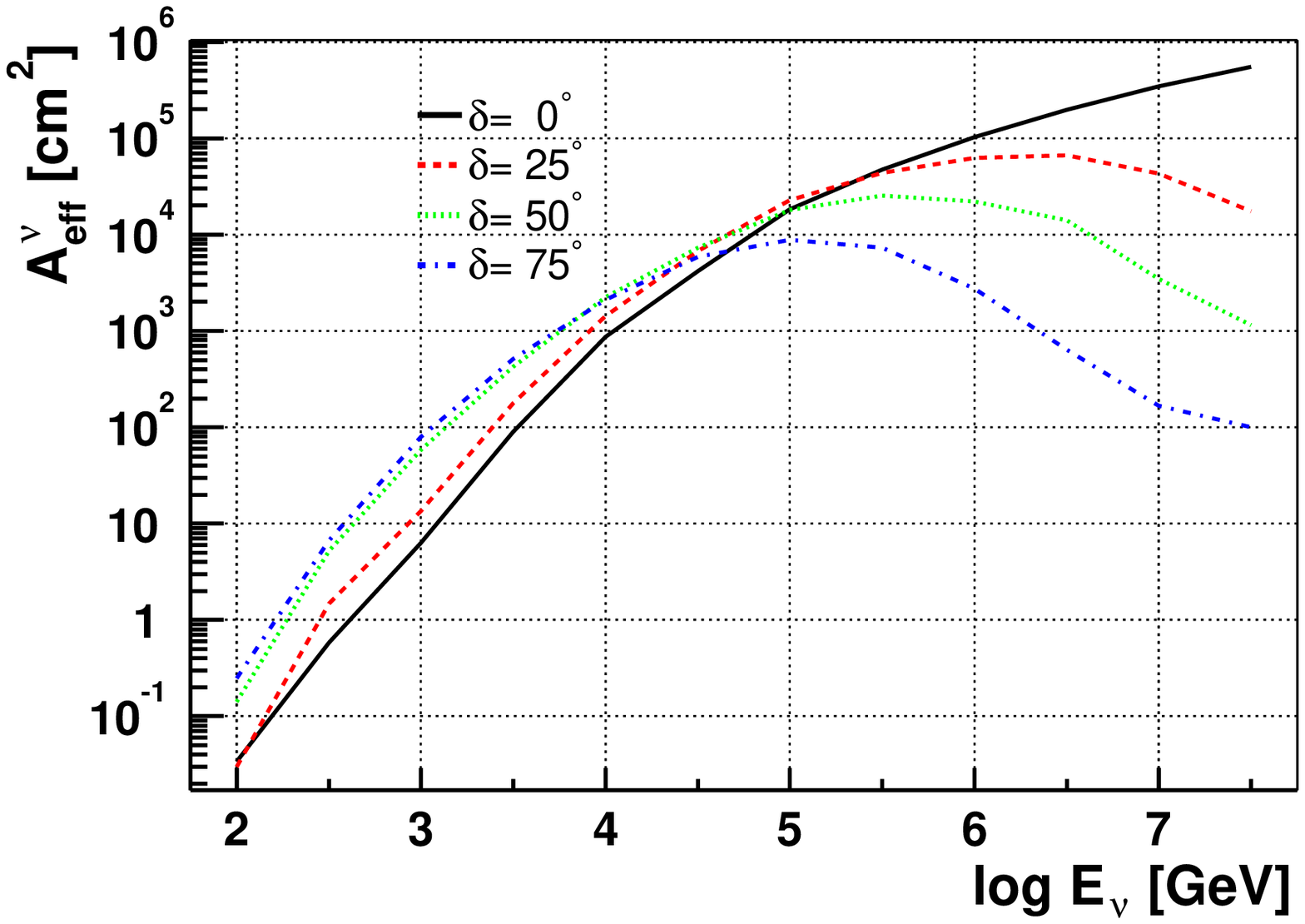} 
\includegraphics[width=3.5in]{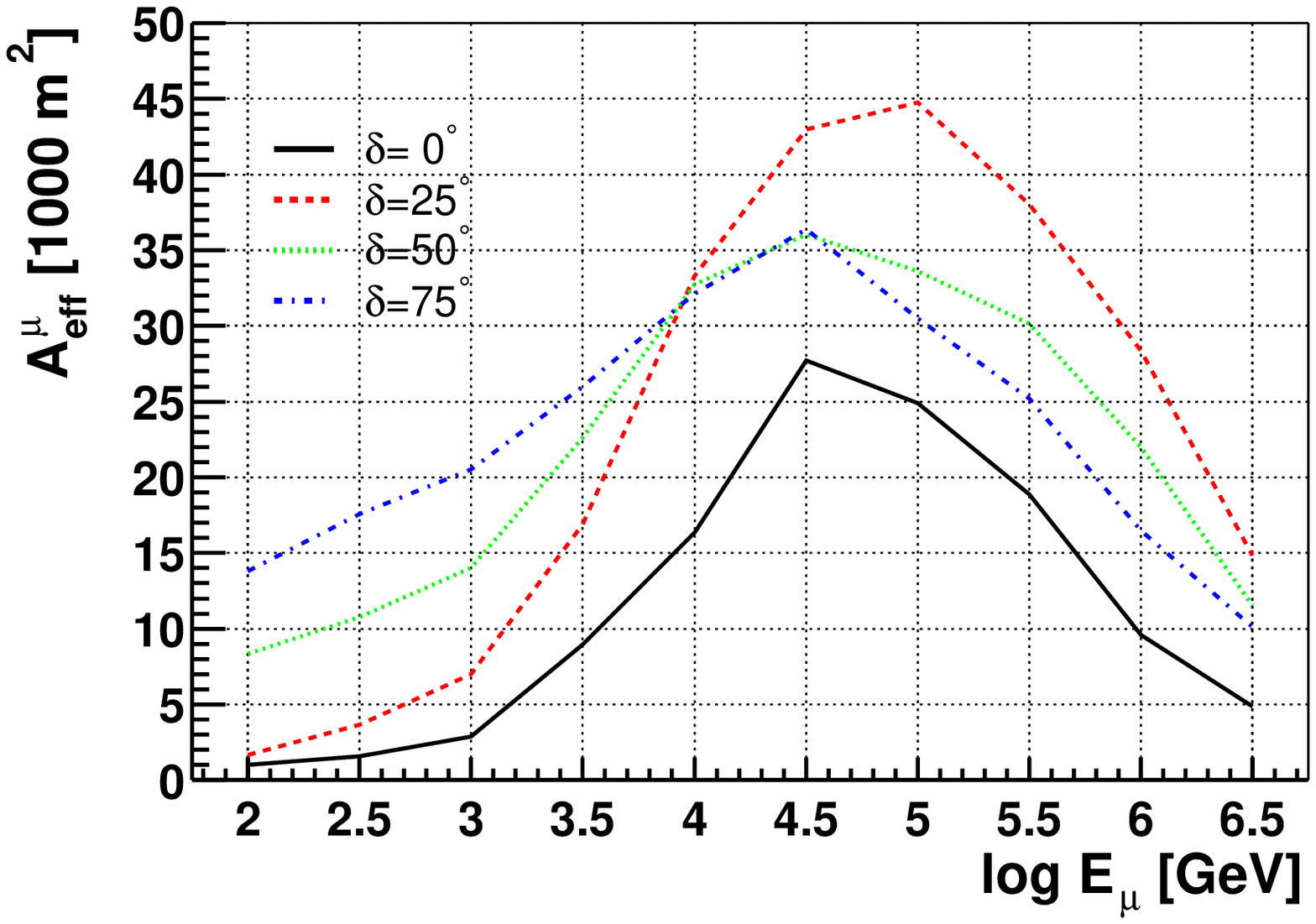}  
\end{center}
\caption{Neutrino and muon effective areas vs energy at different declinations ($\delta$).  
$E_{\mu}$ is the muon energy at closest approach to the center of the detector.
The effect of neutrino absorption in the Earth is included in the neutrino effective areas. }
\label{effarea}
\end{figure*}

In Tab. \ref{agnlimits}, we present neutrino flux limits for northern hemisphere $\mathrm{TeV}$ 
blazars, selected $\mathrm{GeV}$ blazars, microquasars, magnetars, and selected miscellaneous 
candidates. The limits are computed based on an assumed $E^{-2}$ energy spectrum.  
Limits for other spectra can be computed using the neutrino effective area shown in
Fig. \ref{effarea}:
The neutrino flux limit for an assumed flux $d\Phi/dE\propto E^{\alpha}$ is
inversely proportional to the energy averaged effective area
$\overline{\Aeff}(\alpha)=
\int_{E_\mathrm{min}}^\infty \Aeff(E) E^{\alpha} dE
/ \int_{E_\mathrm{min}}^\infty E^{\alpha} dE$,
i.e. a limit for spectral index $\alpha \ne -2$ is obtained by multiplying the presented 
limit by $\overline{\Aeff}(-2)/\overline{\Aeff}(\alpha)$. 
Effective areas at declinations not shown in Fig. \ref{effarea} can be
obtained by linear interpolation in $\delta$;
the systematic shift induced by this interpolation is below 20\% for spectral indices in the range $\alpha=-1.5$\dots$-2.5$.

\begin{table}[htb]
 \begin{center}
  \begin{tabular}{lcccccc}\hline\hline

 Candidate & Dec. [$^\circ$] & R.A. [h] & $n_{\mathrm{obs}}$ & $n_{b}$ & $\Phi_{\mu}^{\mathrm{lim}}$ & $\Phi_{\nu}^{\mathrm{lim}}$ \\ 
   \multicolumn{7}{c}{ \emph{TeV Blazars} } \\
      Markarian 421  & 38.2 & 11.07 &   3 & 1.50 &  3.0 &  3.5 \\
      Markarian 501  & 39.8 & 16.90 &   1 & 1.57 &  1.5 &  1.8 \\   
      1ES 1426+428   & 42.7 & 14.48 &   1 & 1.62 &  1.4 &  1.7 \\
      1ES 2344+514   & 51.7 & 23.78 &   1 & 1.23 &  1.6 &  2.0 \\
      1ES 1959+650   & 65.1 & 20.00 &   0 & 0.93 &  0.9 &  1.3 \\  \hline
   \multicolumn{7}{c}{ \emph{GeV Blazars} } \\
      QSO 0528+134   & 13.4 &  5.52 &   1 & 1.09 &  2.5 &  2.0 \\
      QSO 0235+164   & 16.6 &  2.62 &   1 & 1.49 &  2.0 &  1.7 \\
      QSO 1611+343   & 34.4 & 16.24 &   0 & 1.29 &  0.7 &  0.8 \\
      QSO 1633+382   & 38.2 & 16.59 &   1 & 1.50 &  1.5 &  1.7 \\
      QSO 0219+428   & 42.9 &  2.38 &   1 & 1.63 &  1.4 &  1.6 \\
      QSO 0954+556   & 55.0 &  9.87 &   1 & 1.66 &  1.3 &  1.7 \\
      QSO 0716+714   & 71.3 &  7.36 &   2 & 0.74 &  2.9 &  4.4 \\ \hline
   \multicolumn{7}{c}{ \emph{Microquasars} } \\ 
      SS433          &  5.0 & 19.20 &   0 & 2.38 &  1.0 &  0.7 \\
      GRS 1915+105   & 10.9 & 19.25 &   1 & 0.91 &  2.9 &  2.2 \\
      GRO J0422+32   & 32.9 &  4.36 &   2 & 1.31 &  2.9 &  2.9 \\
      Cygnus X1     & 35.2 & 19.97 &   2 & 1.34 &  2.2 &  2.5 \\
      Cygnus X3     & 41.0 & 20.54 &   3 & 1.69 &  3.0 &  3.5 \\
      XTE J1118+480  & 48.0 & 11.30 &   1 & 0.92 &  1.7 &  2.2 \\
      CI Cam         & 56.0 &  4.33 &   0 & 1.72 &  0.6 &  0.8 \\
      LS I +61 303   & 61.2 &  2.68 &   0 & 0.75 &  1.0 &  1.5 \\  \hline
   \multicolumn{7}{c}{ \emph{SNR, magnetars \& miscellaneous} } \\
      SGR 1900+14    &  9.3 & 19.12 &   0 & 0.97 &  1.4 &  1.0 \\
      Crab Nebula    & 22.0 &  5.58 &   2 & 1.76 &  2.6 &  2.4 \\
      Cassiopeia A   & 58.8 & 23.39 &   0 & 1.01 &  0.9 &  1.2 \\  
      3EG J0450+1105 & 11.4 &  4.82 &   2 & 0.89 &  4.2 &  3.2 \\
      M 87           & 12.4 & 12.51 &   0 & 0.95 &  1.3 &  1.0 \\
      Geminga        & 17.9 &  6.57 &   3 & 1.78 &  3.7 &  3.3 \\ 
      UHE CR Triplet & 20.4 &  1.28 &   2 & 1.84 &  2.4 &  2.3 \\
      NGC 1275       & 41.5 &  3.33 &   1 & 1.72 &  1.4 &  1.6 \\
      Cyg. OB2 region. \cite{ob2} & 41.5 & 20.54 &   3 & 1.72 &  2.9 &  3.5 \\ 
      UHE CR Triplet & 56.9 & 12.32 &   1 & 1.48 &  1.4 &  1.9 \\

 \hline\hline

  \end{tabular}
  \caption{\label{agnlimits}90\% upper limits on candidate sources.  
The number of events observed within the search bin is denoted by 
$n_{\mathrm{obs}}$,
and $n_{b}$ is the number of expected background events determined by 
measuring the background off-source in the same
declination band. Limits are for an assumed $E_{\nu}^{-2}$ spectral shape, 
integrated above $E_{\nu} = 10\, \mathrm{GeV}$, and presented in units of 
$10^{-15} \mathrm{cm}^{-2} \, \mathrm{s}^{-1}\, (\Phi_\mu)$ and $10^{-8} \mathrm{cm}^{-2} \, \mathrm{s}^{-1}\, (\Phi_\nu)$. }
 \end{center}
\end{table} 

In \cite{guetta}, expected numbers of neutrino induced muon 
events for galactic microquasars using source parameters estimated from radio
observations are calculated. For microquasars whose jets have not been resolved in the radio
band the neutrino emission is estimated from the synchrotron luminosity. In the case of the
microquasar SS433, 252 muons $\mathrm{yr}^{-1} \, \mathrm{km}^{-2}$ are predicted. Scaling to 
the AMANDA-II effective area at that declination ($\Aeff^{\mu} = 7900\,\mathrm{m}^{2}$) and to
the livetime of this analysis, yields a prediction of 1.07 events for an assumed $E^{-2}$ spectrum. 
We observe no events in the search bin for this source, and place a 90\% upper limit of 1.24 
events.  Due to a random fluctuation, this limit is about 3 times better than the 
sensitivity at this declination.

\begin{figure}[htb]
\begin{center}
\includegraphics[width=3.2in]{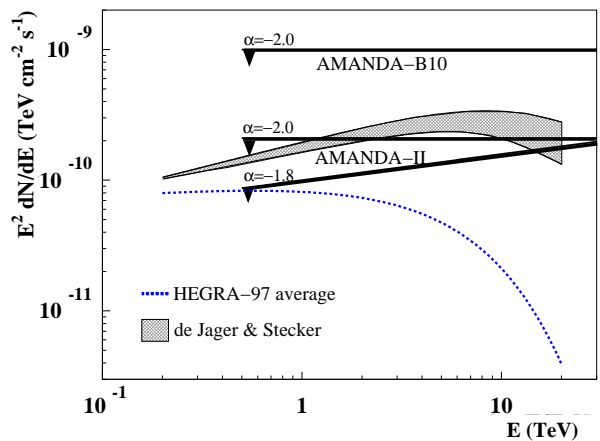}  
\end{center}
\caption{The AMANDA-II 90\% average flux upper limit (197 days livetime) for two assumed spectral indices 
($\alpha$) is compared to the average $\gamma$-ray flux of Markarian 501 as observed in 1997 
by the HEGRA system of air Cherenkov telescopes \cite{hegra97}.
These average upper limits are based on the assumption that the neutrino
spectrum extends to beyond 10 PeV.
Also shown is the 
intrinsic source flux after correction for IR absorption by de Jager and Stecker \cite{dejager}.  
The shaded area is bounded by two curves corresponding to different models of galactic 
luminosity evolution.  For comparison, the AMANDA-B10 result \cite{b10}
is also shown. }
\label{mkn501}
\end{figure} 

In Fig. \ref{mkn501} the AMANDA-II neutrino sensitivity is compared to the 1997-averaged 
TeV $\gamma$-ray flux of the blazar Markarian 501 ($z=0.031$), and the intrinsic source spectrum 
(corrected for IR absorption).  
The figure demonstrates AMANDA-II has achieved the sensitivity needed 
to search for neutrino fluxes from TeV $\gamma$-ray sources of similar strength to the intrinsic $\gamma$-ray flux.

Data collected in 2001-2002
are being analyzed and should improve the sensitivity of this analysis approximately by a factor 2.3.

\section{Acknowledgments}

We acknowledge the support of the following agencies: National
Science Foundation--Office of Polar Programs, National Science
Foundation--Physics Division, University of Wisconsin Alumni Research
Foundation, Department of Energy, and National Energy Research
Scientific Computing Center (supported by the Office of Energy
Research of the Department of Energy), UC-Irvine AENEAS Supercomputer
Facility, USA; Swedish Research Council, Swedish Polar Research
Secretariat, and Knut and Alice Wallenberg Foundation, Sweden; German
Ministry for Education and Research, Deutsche Forschungsgemeinschaft
(DFG), Germany; Fund for Scientific Research (FNRS-FWO), Flanders
Institute to encourage scientific and technological research in
industry (IWT), and Belgian Federal Office for Scientific, Technical
and Cultural affairs (OSTC), Belgium; Fundaci\'{o}n Venezolana de 
Promoci\'{o}n al Investigador (FVPI), Venezuela;  D.F.C. acknowledges 
the support of the NSF CAREER program.

\end{document}